# Enhanced band edge luminescence of ZnO nanorods after surface passivation with ZnS

Asad Ali[a,b*], Gul Rahman[c], Tahir Ali[d], M. Nadeem[d], S.K.Hasanain[e], M.Sultan[*b]

(a) Preston Institute of Nanoscience and Technology, Preston University, Islamabad, Pakistan
(b) Nanoscience and Technology Department, National Centre for Physics, Quaid-I-Azam University Campus, Islamabad, Pakistan
(c) Department of Physics, Quaid-i-Azam university, Islamabad 45320, Pakistan
(d) Physics Division, PINSTECH, Nilore, Islamabad, Pakistan
(e) COMSTECH Secretariat, 33 Constitution Avenue, G-5/2, Islamabad

Correspondence: sultan@ncp.edu.pk    asadqau88@gmail.com

**Keywords:** Hybrid nanostructures, ZnO nanorods, photovoltaics, Photoluminescence

## Abstract

We report on the passivation of surface defects of ZnO nanorods by surface layer deposition. ZnO nanorods and ZnS-ZnO hybrid nanostructures are grown on FTO coated glass substrate by chemical bath deposition method. XRD spectrum of ZnO nanorods shows the preferential growth along the $c$-axis. SEM analysis confirms the nearly aligned growth of the ZnO nanorods with a hexagonal shape. XPS measurements were performed to confirm the deposition of the surface layer and surface stoichiometry. Room temperature photoluminescence of ZnO nanorods showed two emission bands, viz. the band edge emission and the blue-green emission, with the latter being associated with the defect states arising from the surface of ZnO nanorods. The band edge emission is significantly increased as compared to blue-green emission after ZnS surface layer deposition on ZnO nanorods. The quenching of blue-green emission is explained in terms of reduced surface defects after ZnS deposition. Density functional theory (DFT) calculations are used to understand the mechanisms of defect passivation in ZnS-ZnO nanostructures and we show that the S atoms prefer the O site as compared with the Zn and interstitial sites.

## 1. Introduction

ZnO nanostructures have attracted considerable attention due to their excellent electrical and optical properties. These properties make ZnO one of the suitable materials for use in next-generation optoelectronic devices [1]. ZnO is a nontoxic material and can be synthesized using cost-effective solution process such as Chemical Bath Deposition (CBD) method. ZnO based nanostructures have wide range of applications in a device technology. They have been used in photovoltaics in different device architectures such as in planar and rod like geometries. ZnO nanostructures are also been used in gas and bio sensors, transistors, optically pumped lasers, light emitting diodes, and piezoelectric devices etc [2-5]. The wide band gap (3.3 eV) and high exciton binding energy (60 meV) at room temperature make ZnO one of the most extensively used semiconductor materials [6]. In recent years, There has been a surge in the investigation of Photoluminescence of ZnO nanostructures [7]. In most of the studies, two emission bands have been observed [8] such as the near band emission which corresponds to band-to-band transition of the charge carriers and the defect emission which is the result of structural defects of ZnO nanostructures, such as Zn interstitials, oxygen vacancies, dangling bonds at the surface etc. These structural inhomogeneities result in the formation of inter band states. Optoelectronic properties of nanomaterials are greatly affected by the confinement effects and surface defects [9-11]. The latter can cause the defect emission and band bending which results in the low luminescence efficiency of the nanomaterials [12-13]. Many researchers attributed the defect emission band in PL of ZnO nanostructures to the surface states [14], while others have argued that the green emission might be the result of bulk defects, such as oxygen vacancies [15]. Suppressing the surface defects to enhance the ultraviolet (UV) emission of ZnO nanorods can result in better performance of the ZnO based optoelectronic devices.

There are few reports in which the optical properties of ZnO nanostructures have been altered by surface treatments [15-17]. Richters et al. have investigated the PL properties of $Al_2O_3$/ZnO nanowire structures [15] where they have found that after coating the ZnO nanowires with $Al_2O_3$ the near band emission at low temperatures was enhanced and the deep level emission was reduced. It has also been observed that hydrogenation also results in reduced deep level emission and increased band edge emission [17]. Other surface treatments such as argon ion milling and polymer covering have also been observed to reduced deep level emission and produce increased band edge emission in ZnO nanostructures [18-19]. The reduced defect emission is expected to result in better performance of the ZnO based hybrid nanostructures.

Considering the good electronic transport properties of ZnO, it has been used as an electron transport material in hybrid solar devices. However, utilization of the ZnO nanostructures in modern solar cell devices has remained limited due to its surface defects which not only contribute to reduction of Power Conversion Efficiency (PCE) but also lead to degradation of absorber materials [20-21]. As discussed above that one of the possible ways to reduce defects in ZnO nanostructures is their surface treatments. From the device point of view one needs to select a material for the surface treatment that offers a suitable band alignment in hybrid nanostructures to facilitate charge transport as well as serve to reduce surface defects in ZnO nanostructures. In this study, we investigate the effect of ZnS surface layer on the photoluminescence and electrical properties of ZnO nanorods, and find substantial enhancement in band-edge emission and suppression of the defect emissions after surface treatment.

## 2. Experimental Section

### 2.1. Growth of Zinc Oxide Nanorods

We first washed the FTO substrates in the baths of acetone, methanol, ethanol, and in distilled water sequentially using sonicator. After washing the substrates, we deposited ZnO seed layer using spin coating method. ZnO seed layer solution was prepared using Zinc acetate hexahydarte, ethylene glycol and diethanol amine. First we prepared 0.4 M solution of Zinc acetate in ethylene glycol. We noticed that the solution turned milky after few minute stirring at room temperature [21]. We added Diethonl ammine drop wise in the solution of Zinc acetate until it becomes transparent. We stirred the solution further for 30 minutes and the final solution was spin coated on FTO glass at 2500 rpm for 30 sec. After each spin coat the samples were dried in oven at 250 $^0$C for 20 minutes. We repeated this process five times and the samples were annealed at 500 $^0$C for two hours [22].

ZnO nanorods were grown using Chemical Bath Deposition method (CBD). The growth solution was prepared using aqueous solution of zinc nitrate and hexamethylene tetramine (HMTA). First we separately prepared the equimolar aqueous solution of HMTA and Zinc nitrate hexahydrate. After 30 minutes stirring, we pour the solution of HMTA into zinc nitrate hexahydrate solution and the final obtained solution was stirred further for one hour. Substrates were immersed in the growth solution with the support from the Teflon rod, such that the seed layer coated side of the substrates were facing down. The reaction time for the growth of ZnO nanorods was 2 hours at 90 $^0$C. After the completion of reaction, we washed the samples with distilled water and finally annealed at 400 $^0$C for one hour [23-24].

### 2.2. Growth of ZnS-ZnO Nanostructures

For the growth of ZnS shell on ZnO nanorods, we used simple chemical conversion route. We prepared 0.2 M aqueous solution of thioacetamide (TAA) and the substrate with ZnO nanorods array was immersed in this solution. The sulfidation of ZnO NRs takes place due to the large difference between the solubility product constant ($K_{sp}$) between ZnO and ZnS [25]. The value of $K_{sp}$ for ZnO is $6.8 \times 10^{-17}$ and for ZnS, its value is $2.93 \times 10^{-25}$ which makes a practical way for the conversion of ZnO into ZnS. During the sulfidation process, TAA hydrolyzes and formation of $H_2S$ gas takes place. The $S^{2-}$ reacts with the whole surface of ZnO nanorods and the formation of ZnS shell takes place. We carried out the sulfidation process at 90 °C for 15 minutes. The samples were finally washed with distilled water and dried at room temperature. [26-27].

**2.3. Characterization Methods**

Different characterization techniques were used for the detailed analysis of the samples. XRD analysis was used to confirm the crystal structure of ZnO nanorods and ZnS nanostructures (D8 diffractometer, Cu Kα λ= 0.1541nm). Hitachi S4800 field emission microscope was used to study the surface morphology as well as EDX analysis of the samples. Lambda 950 (Perkin Elmer) spectrometer (UV/Vis/NIR) was used for the measurement of optical transmittance of the samples. For XPS Measurements we used standard omicron system which is equipped with monochromatic Al Kα X-ray source (1486.7 eV) and Argus hemispherical electron spectrometer with 128 channels MCP detector. We made the measurements in ultra high vacuum conditions. The samples were Air expose before the XPS measurements. For XPS data analysis we used CASA XPS software and used C1s for the calibration of the binding energy. For solid-state impedance analysis, Alphas-N (Novocontrol Germany) was employed in the frequency range of 1 Hz to 10 MHz at room temperature. Connections were made on both sides of the nanostructures and a 0.5 V ac signal was used for the impedance study.

Density functional theory (DFT) calculations with local density approximation (LDA) [28] using linear combination of atomic orbital basis as implemented in the SIESTA code [29] were performed to study different kinds of defects in ZnO nanorods. A double-ζ polarized basis set for all atoms was used. We used standard norm-conserving pseudopotentials [30] in their fully nonlocal form. A cutoff energy of 200 Ry for the real-space grid was adopted. Atomic positions were relaxed, using conjugate-gradient algorithm [31], until the residual Hellmann-Feynman force on a single atom converges to less then 0.05 eV/Å.

**3. Results and Discussion**

**3.1 Structural Characterization**

The pattern (a) in Figure 1 represents the XRD spectrum of ZnO nanorods grown on FTO coated glass substrate. This spectrum shows diffraction peaks at different 2-theta values. We can label

the peaks as (100), (002), (101), (102), (103) and (004) at 31.80$^0$, 34.40$^0$, 36.30$^0$, 47.60$^0$, 630$^0$ and 72.80$^0$, respectively and all the peaks matched with the previously published report [32]. These observed planes confirm the hexagonal wurtzite type structure of ZnO with space group *P63mc*. The intensity of (002) peak is quite sharp indicating the orientation of ZnO nanorods along the *c*-axis and perpendicular to the FTO Substrate. The XRD pattern indicates that the sample obtained by the aforementioned CBD technique is comprised of only ZnO while no other Zn or O containing phase is observed. However, there are some additional reflections (marked by asterisks) in the XRD pattern which are contributed by the FTO substrate. The pattern (b) in Fig.1 shows the XRD spectrum of ZnS-ZnO hybrid nanostructure. In addition to FTO and ZnO peaks, a hump at 28.5$^o$ is observed which corresponds to the (002) plane of hexagonal phase ZnS (JCPDS-36-1450) [33]. The inset of Fig.1 shows the magnified XRD pattern of ZnS-ZnO nanostructures. In this pattern we can clearly see a broad hump at 28.5$^0$. The emergence of (002) plane shows that the growth of ZnS layer has taken place on the surface of ZnO nanorods.

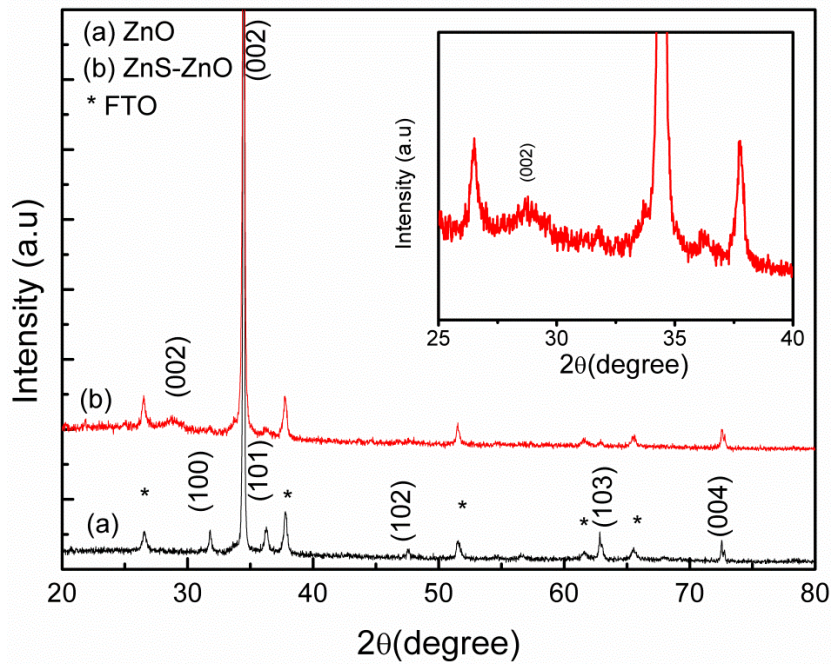

**Figure 1: (a) XRD of ZnO nanorods on FTO substrate. (b) XRD of ZnS-ZnO hybrid nanostructures, Inset: Magnified XRD pattern of ZnS-ZnO.**

### 3.2 SEM and EDX Analysis

Figure 2 presents the SEM micrographs of ZnO nanorods and ZnS-ZnO hybrid system at different resolutions. Figure 2(a) and 2(b) present the SEM top view of ZnO nanorods grown on FTO substrate, while Fig. 2(c) and 2(d) show SEM micrographs of ZnS-ZnO nanostructure.

SEM micrographs confirm the hexagonal morphology of the grown ZnO nanorods. The average diameter of the nanorods is estimated between 80 nm to 90 nm. After growing ZnS layer on the surface of the ZnO nanorods the hexagonal morphology of the nanorods stays the same which shows that the chemical method used for the growth of ZnS shell has no effect on the overall morphology of the nanorods. We do not see a sizeable change in the diameter of the ZnO nanorods after the growth of the ZnS surface layer. As mentioned in the experimental section, in this method the ZnS layer grows on the surface of ZnO nanorods by replacing the O of the structure with S, therefore no substantial change is expected in the diameter of the nanorods.

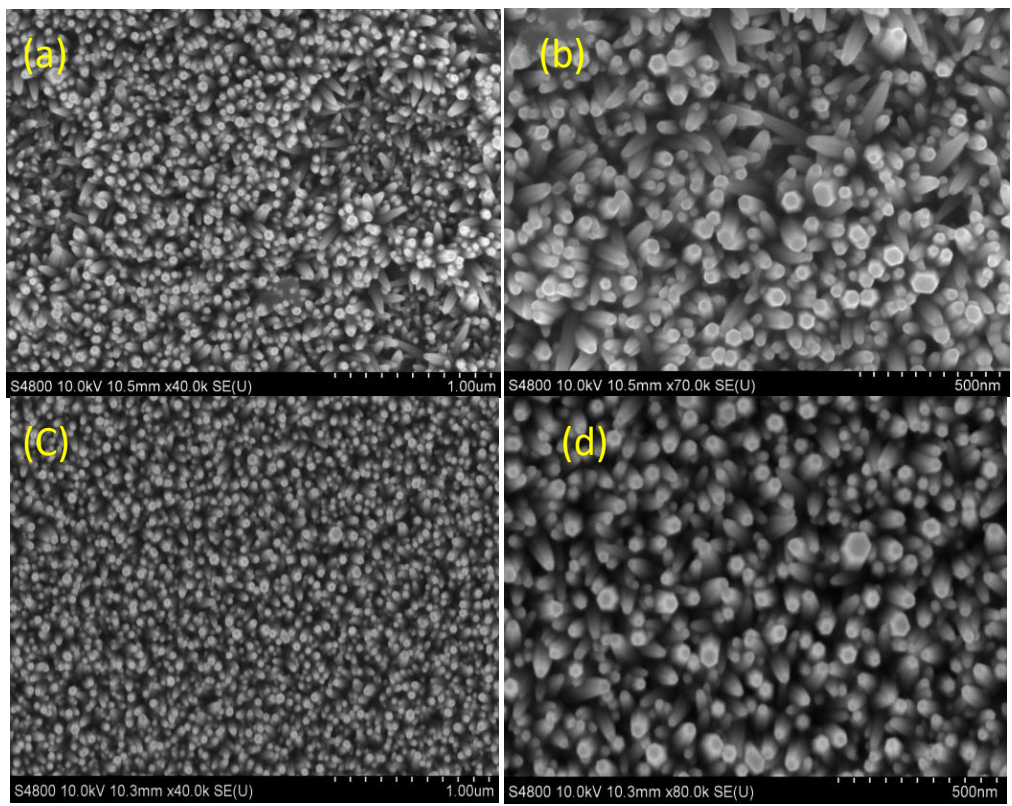

**Figure 2: (a) and (b) SEM micrographs of ZnO nanorods on FTO glass. (c) and (d) SEM micrographs of ZnS-ZnO hybrid nanostructures**

To get the qualitative analysis of the composition of the grown ZnO nanorods and ZnS-ZnO hybrid nanostructures, we performed EDX of our samples. Figure 3 (a) shows the EDX spectra of ZnO nanorods with major peaks coming from Zn and O while no other impurities are detected. The EDX spectrum of ZnS-ZnO hybrid nanostructures is exactly similar (Not shown here) to ZnO nanorods and showed only peaks from Zn and O, while no sulphur has been detected which indicates that the thickness of ZnS layer is very thin. The XRD spectrum shows a very weak peak corresponding to ZnS and the XPS discussion in the latter section confirms that the thickness of ZnS is less than photolectrons's penetration depth (> 8 nm). Due to less thickness of ZnS the excitation volume for the generation of X-rays would be less as a result we did not observe any peak related to sulphur in the EDX spectrum.

To confirm the elemental distribution, we also performed EDX X-ray elemental mapping of the samples. Figure 3 (b-c) shows the mapping images of ZnO nanorods. It shows a uniform distribution of Zn and O elements. The EDX mapping images of ZnS-ZnO sample had almost the same distribution of Zn and O elements as like ZnO nanorods. In this sample, in addition to Zn and O, we also observed sulphur. Figure 3-d shows the EDX mapping image of the S and confirms its uniform distribution. It is clear from the mapping image that the amount of sulphur is less as a result there will be low content of ZnS and consistent with our earlier discussed XRD and EDX results.

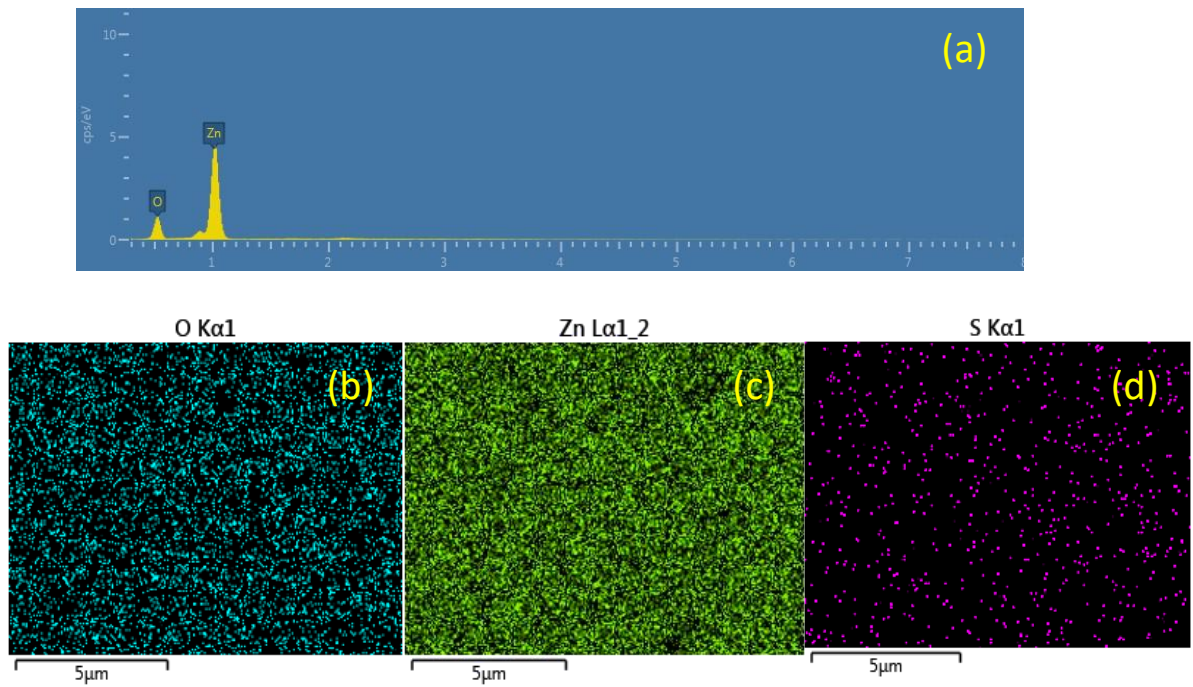

**Figure 3 (a): EDX spectrum of ZnO nanorods (b-d) EDX X-ray elemental mapping images of O, Zn, and S**

### 3.3 XPS Analysis

Figure 4 presents the XPS spectra of ZnO nanorods and ZnS-ZnO hybrid nanostructures. Oxygen 1s spectrum presented in Fig. 4(a) can be deconvoluted into three components at the binding energies of 530.2 eV, 531.8 eV, and 532.8 eV. The peak at the low binding energy corresponds to the lattice oxygen peak where the $O^{-2}$ ions in the structure of ZnO are surrounded by the Zn atoms. The other peaks at the binding energy of 531.8 eV and 532.8 eV are originating from the surface adsorbed species such as chemisorbed or dissociated oxygen or OH species on the surface of ZnO nanorods [21]. Some researchers also assign these binding energy contribution around 531.8 eV to the oxygen vacancies at the surface of ZnO [34].Here, it is difficult to separate the oxygen vacancies peak from the surface adsorbed species as all the three samples are air exposed and we expect the major contributions in the higher binding energies are originating from the surface adsorbed species. In order to confirm the origin of this peak, we grew a thicker layer of ZnS on ZnO nanorods by increasing the growth time of ZnS layer up to 30 minutes. Fig. 4(a) (Top section) presents the O1s XPS spectra for the thicker ZnS layer grown on ZnO nanorods shows the diminishing oxygen lattice peak which suggests that the thickness of the ZnS layer is actually exceeding the photoelectron's penetration depth. It is worthwhile to note that the contribution of the O1s peak at the higher binding energy is almost the same as for the bare ZnO nanorods and for the sample with ZnS layer grown by 15 minutes reaction time. The same trend of these two higher binding energy peaks in all the three samples suggest that the surface adsorbed contribution on all surfaces is similar since all samples had similar air exposure.

Figure 4(b) shows the Zn 2p core levels of ZnO nanorods and ZnS-ZnO hybrid nanostructure. The observed binding energy of Zn $2p_{3/2}$ in case of ZnO nanorods is 1021.6 eV while in case of ZnS-ZnO hybrid system the observed binding energy of the same level is 1021.9 eV. There is a shift in binding energy of about 0.3 eV. This shift in the binding energy shows that Zn is present in the form of ZnS at the outer surface of ZnO nanorods and confirms the presence of ZnS. Our DFT calculations (discussed below) also show that S can easily be doped at the O site in ZnO, which can form ZnS bonds.

Figure 4 (b) (top) also shows the sulphur 2s peak for ZnS-ZnO hybrid system. This peak can be deconvoluted into two components at the binding energy of 162.0 eV and 163.2 eV. The component at the binding energy of 162.0 eV represents sulphur $2p_{3/2}$ while the component at 163.0 eV represents sulphur $2p_{1/2}$. Since we observed only one group of sulfur 2p peaks for ZnS-ZnO hybrid nanostructures, it confirms that there is only one valence state for the sulfur corresponding to $S^{-2}$ in ZnS.

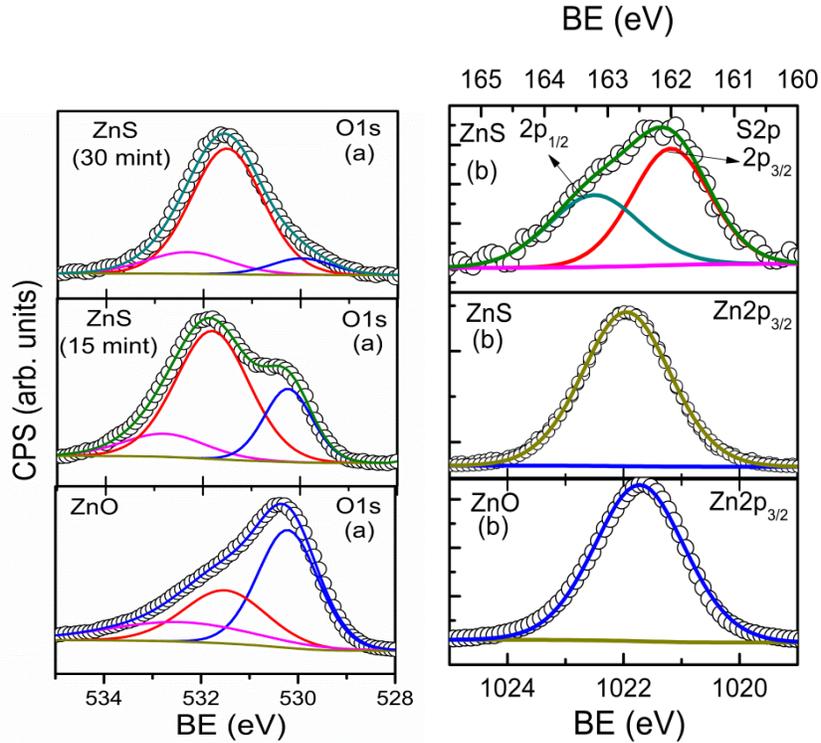

**Figure 4: (a)** Oxygen 1s XPS spectra taken from bare ZnO nanorods and ZnS-ZnO hybrid nanostructures grown for different times. **(b)** XPS spectra of Zn 2p core level taken from ZnO and ZnS-ZnO (bottom), S2p core level spectrum of ZnS-ZnO hybrid nanostructure (Top)

### 3.4 Optical Properties

To investigate the optical properties of the ZnO nanorods and ZnS-ZnO hybrid nanostructures, we performed the room temperature photoluminescence and optical transmittance of our samples. Figure 5 presents the room temperature photoluminescence of ZnO nanorods and ZnS-ZnO nanostructures. In this figure we can see two emitting bands i.e. a narrow ultraviolet emission peak which is centered around 378 nm along with a broad blue-green defect emission for both ZnO nanorods and ZnS-ZnO hybrid nanostructure. It is evident that the band edge emission increases in ZnS treated sample as compared with bare ZnO nanorods. On the other hand the broad blue-green emission decreases substantially with the treatment. When the bare ZnO nanorods sample is excited by the laser source and its electrons are excited to the conduction band, only a few of these electrons relax back directly to the valence band while the majority of them are trapped in the defect levels. Thus the electron from the conduction band is readily trapped back into the defect states to produce a broad hump in the visible region.

The Gaussian fit (Fig. 5(b)) of the PL data shows multiple peaks both in UV and in visible range. In UV region the excitonic peaks at 378 nm and 385 nm relate to the first and the second order phonon replica, respectively, of a free exciton band in hexagonal wurtzite structure [35]. In the visible region the blue-green luminescence in ZnO nanostructures has been associated with different types of defects such as oxygen vacancies ($V_o$) [36-37], zinc interstitials ($Zn_i$) [38], oxygen interstitials ($O_i$), zinc vacancies ($V_{zn}$) and antisite oxygen ($O_{zn}$) [39]. Multiple reports have shown that by decreasing the surface to volume ratio in ZnO nanostructures the intensity of green emission decreases [40-44]. We also performed DFT calculations which have been discussed in the latter section and have supported this fact. So we expect that the observed defect emission is mostly contributed by the surface defects.

First, we address the origin of the green emission in our experimental PL result. Vlasenko et al. [45], Ahn et al.[46] and Cao et al. [47] disentangled two possible contributions to the green luminescence at ~554 nm and ~504 nm by using electron paramagnetic resonance. Both of their proposed contributions has also been observed in our result and fits very well with their findings. They have suggested that the component at ~ 554 nm is due to the transition from $Zn_i$ to $V_o$ while the component at ~504 nm results from the transition between CB to $V_o$. The component observed at ~ 519 nm has been associated with the oxygen antisite ($O_{zn}$) defect. $O_{zn}$ level is located approximately 0.1 eV above the valence band and the transition from the conduction band to the $O_{zn}$ defect level results in the emission of 519 nm component [39]. The component at ~ 541 nm has been associated with the $O_i$ defects [48]. $O_i$ defects result in the formation of shallow defect level at about 1.08 eV above the VB. The transition from the conduction band to the $O_i$ defect states results in the emission of 541 nm component.

The blue luminescence in ZnO nanostructures has been associated with $Zn_i$ and $V_{zn}$ defects in ZnO nanostructures. The transition from the extended $Zn_i$ defects to the valance band results in the emission of components at ~446 nm and ~468 nm [49]. The component at ~487 nm has been associated with the transition from the $Zn_i$ level closer to the conduction band, to the Zn vacancy levels which are formed near the valence band [50-51].

It is important to note that after growing ZnS layer on the surface of ZnO nanorods the intensity of the defects emission reduces significantly. There could be two different phenomena for the enhanced UV emission and the reduced defect emission. ZnS has a larger band gap (3.7 eV) as compared to ZnO (3.3 eV) ; as a result the confinement of the charge carriers will take place in ZnO and hinders the tunnelling of charges from ZnO to ZnS outer surface. There will be more photogenerated electron-hole pairs in ZnO which may result in enhanced UV emission and reduced defect emission [52]. Another possible mechanism for the reduced defect emission and the enhanced band edge emission is the surface defect passivation after growing ZnS surface layer. As discussed above there are different types of defects which are contributing to the emission. The possible reason for the reduced defect emission related to oxygen vacancies is filling of the vacant oxygen sites by sulfur atoms. Both oxygen and sulfur are isoelectronic while sulfur can also exist in the -2.0 oxidation state; hence there is a high probability of filling the vacant oxygen sites by the sulfur atoms to form ZnS.

The method that we have used for the growth of ZnS is the actual conversion of ZnO nanorods itself into ZnS by dissolving oxygen of ZnO and replacing it with by sulfur atoms. We expect that oxygen related defects may have been reduced during the sulfurization process of ZnO nanorods. There are three different sites viz. $O_i$, $O_{zn}$ and lattice oxygen where the introduced sulfur atoms can enter and replace the dissolved oxygen. By dissolving, some of the $O_i$ and $O_{zn}$ in the structure of ZnO will ultimately result in reduced mid-gap states and could be the possible reason for the reduced defect emission related to these defects. The $Zn_i$ related defects may have been reduced due to the bonding of S with the $Zn_i$ atoms during the sulfrization process to form ZnS. The octahedral site is the most favorable position for the $Zn_i$ atoms and we expect that the sulphur may have formed bonding with these interstitial Zn atoms to form ZnS which is the possible stable state.

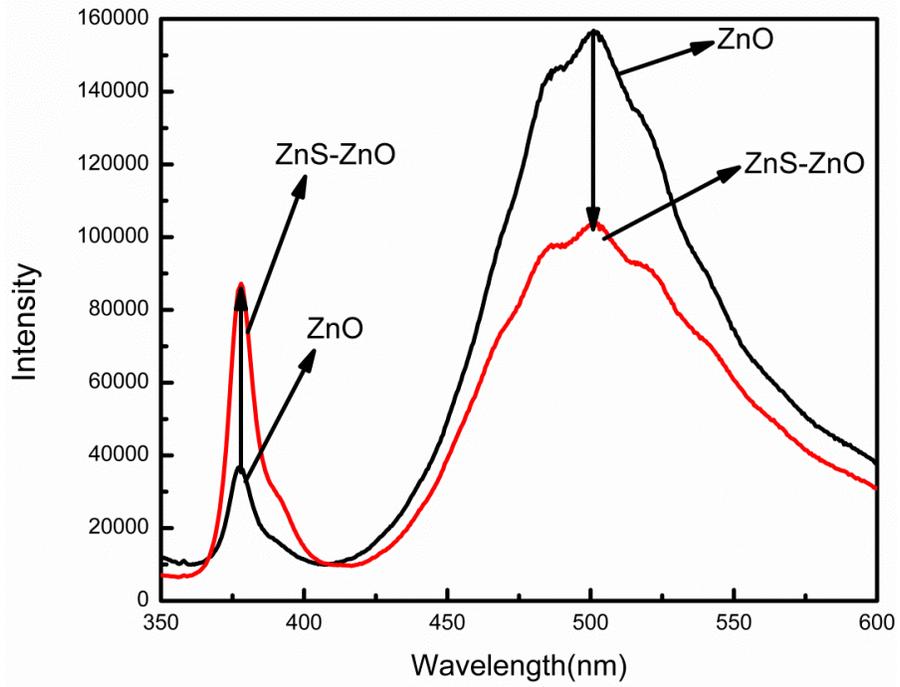

**Figure 5: Room temperature Photoluminescence of ZnO nanorods and ZnS-ZnO nanostructures**

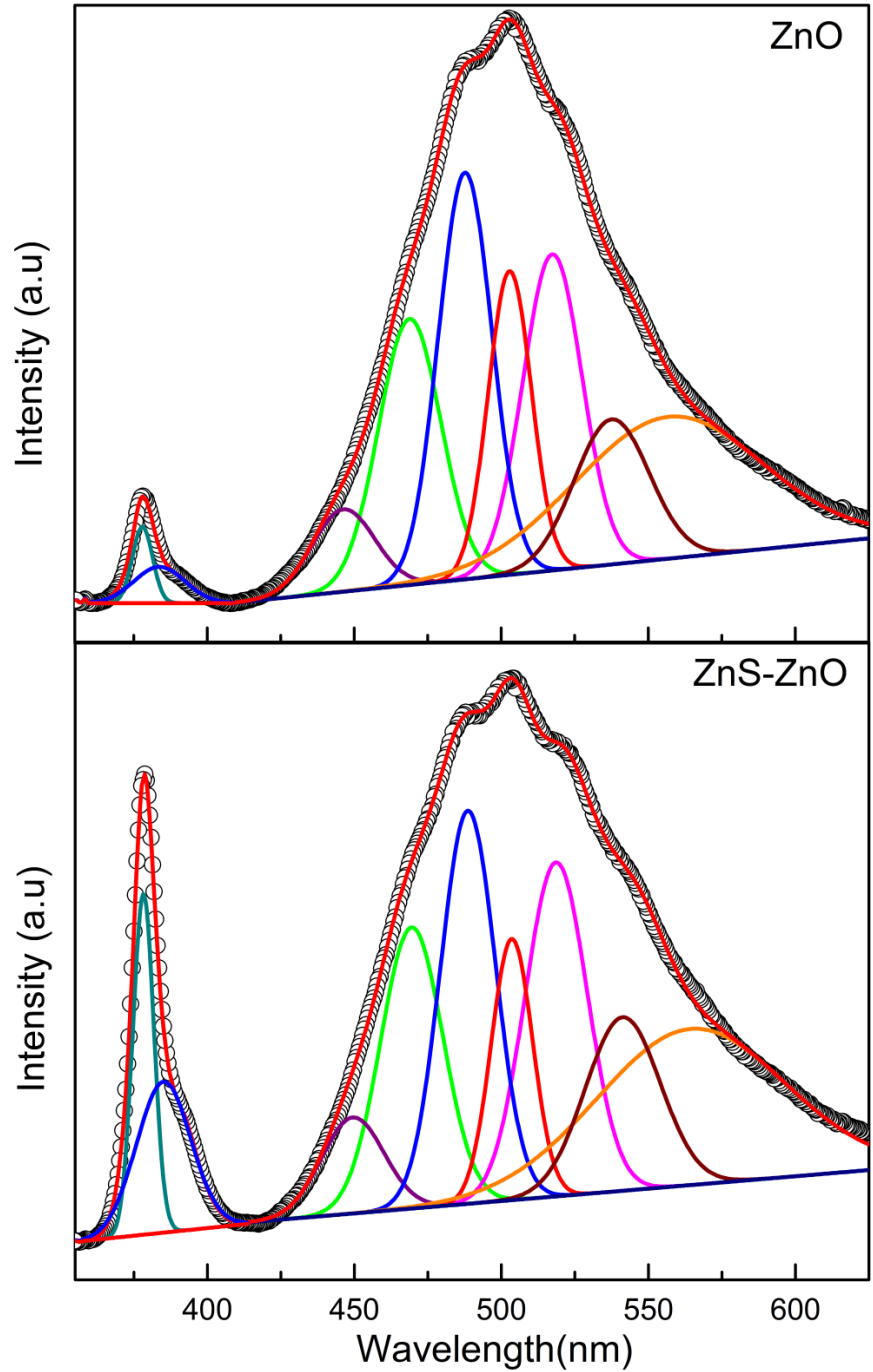

**Figure 5 (b): Gaussian fitting of multiple peaks in the PL data**

Figure 6 presents the optical transmittance of FTO/ ZnO NRs and FTO/ZnO NRs / ZnS nanostructures. Both the samples showed high transmittance in the visible range due to their wide bandgap. In case of ZnO nanorods, there is a sharp absorption edge at about 378nm which corresponds to the band gap of ZnO. The band gap value was determined from the experimental reflectance data using the Kabulka Munk function and the calculated value of the band gap is 3.28 eV. In case of ZnS-ZnO sample, there is an extra absorption edge at about 335nm in addition to the band edge of ZnO. This extra absorption edge corresponds to the ZnS and confirms the growth of ZnS layer on ZnO nanorods. The ZnO and ZnS-ZnO optical transmittance spectra closely resembles previously published report [21]. Following the above PL discussion, the optical transmittance data are also showing some signs of intrinsic defects in ZnO nanorods. If we compare the optical transmittance spectra of both ZnO and ZnS-ZnO samples, we observe a slight increase in transmittance of ZnS-ZnO hybrid nanostructure as compared to ZnO nanorods in the visible range. This shows that the defect states of ZnO nanorods are absorbing some light in the visible range but after the growth of the ZnS shell some of the defect states were passivated and as a result the ZnS-ZnO sample showed increased transmittance in the visible range. The optical transmittance data are also supporting the above discussed PL result and confirm that ZnS layer is helping in passivation of surface defects of ZnO nanorods.

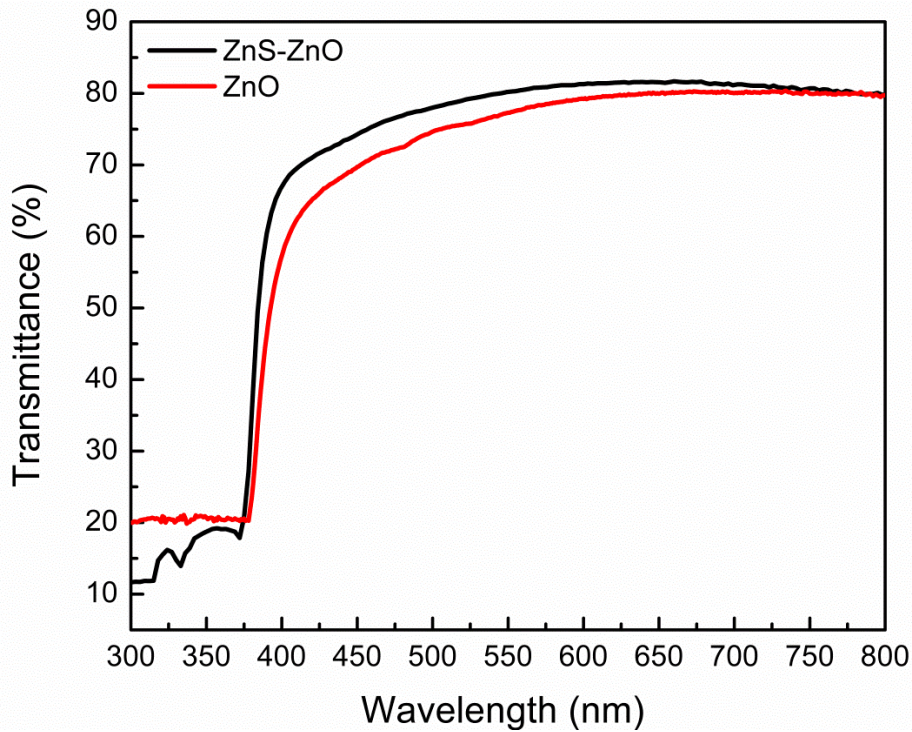

**Figure 6: Optical Transmittance of FTO-ZnO and FTO-ZnS-ZnO hybrid nanostructures**

**3.5 Impedance Spectroscopy**
AC electrical response in polycrystalline materials generally elucidate the effects of electrode-semiconductor contact, inter-grain surface (grain boundaries) and intra-grain effects at low, intermediate and high frequencies, respectively [53]. Figure 7 shows; impedance plane plot of ZnO nanorods is larger than that of ZnS shell coated on ZnO nanorods. The semi-circular arc intersects at the right-hand side of real part of impedance and gives the resistance corresponding to dc values ($R_{dc}$ as frequency tends to zero). The extension of a semicircle on left-hand side intersects the x-axis at the point labeled as $R_1$, the contact resistance or the bulk resistance whose arc is not complete due to a limitation in the applied frequency range. The variation in the magnitude of the semi-circular arc at the expense of defects and structural stresses can be addressed from the term depression angle using ZView software. The depression angle is defined as the displacement of impedance semi-circles below the real axis due to the presence of distributed elements in the matrix [54]. Depression angles calculated for ZnO nanorods and ZnS-ZnO hybrid nanostructures are $1.7^0$ and $6.6^0$, respectively.

Using simulated fit circle model, following parameters are calculated; $R_1 = 28$ ohms, $R_2 = 300$ ohms and capacitance $C_2 = 6.37 \times 10^{-10}$ F. $R_1$ has been defined above while $R_2$ is the resistance value of electro-active region defined by semi-circular arc and $C_2$ is the capacitance. An equivalent circuit model $R_1(R_2C_2)$ is employed to fit impedance plane plots and calculated values of $R_1 = 30$ ohms, $R_2 = 297$ ohms and $C_2 = 6.51 \times 10^{-10}$ F. For ZnS-ZnO nanostructures simulated equivalent circuit $R_1(R_2C_2)$ gave values as $R_1 = 14.9$ ohms, $R_2 = 173$ Ohms and $C_2 = 8.03 \times 10^{-10}$ F. Using $R_1(R_2C_2)$ equivalent circuit model, fitting is carried out but the fit was not good. In view of higher depression angle value of ZnS-ZnO, constant phase element is introduced instead of capacitance to compensate the presence of heterogeneity in the sample. An equivalent circuit model $R_1(R_2Q_2)$ is employed to fit ZnS-ZnO sample, where Q is the constant phase element and is used to compensate non-ideal behavior of the capacitance. Using fit model, calculated values of fitting parameters are; $R_1 = 14.8$ ohms, $R_2 = 172$ ohms and $Q = 2.51 \times 10^{-10}$ F and $n = 0.93$. For a pure capacitor, $n = 1$ and lower value of n for ZnS-ZnO indicates deviation from ideal capacitor behavior. The low value of n can be explained due to the inhomogeneity introduced by ZnS core at ZnO. The change in the impedance of ZnS-ZnO is further explained in the log-log plot conductivity with frequency. Figure 7(b) shows a visible change in the frequency independent part and these results are in accordance with early discussed impedance results.

Introduction of ZnS layer to ZnO nanorods results in more effective charge separation and is interpreted as a faster interfacial charge transfer. These results are also in accordance with the PL results where spectra of ZnO-ZnS show a diminishing trend in the defected related emissions.

The high value of depression angle of ZnS-ZnO hybrid nanostructures is obvious due to the structural stress created by coated ZnS layer on ZnO nanorods.

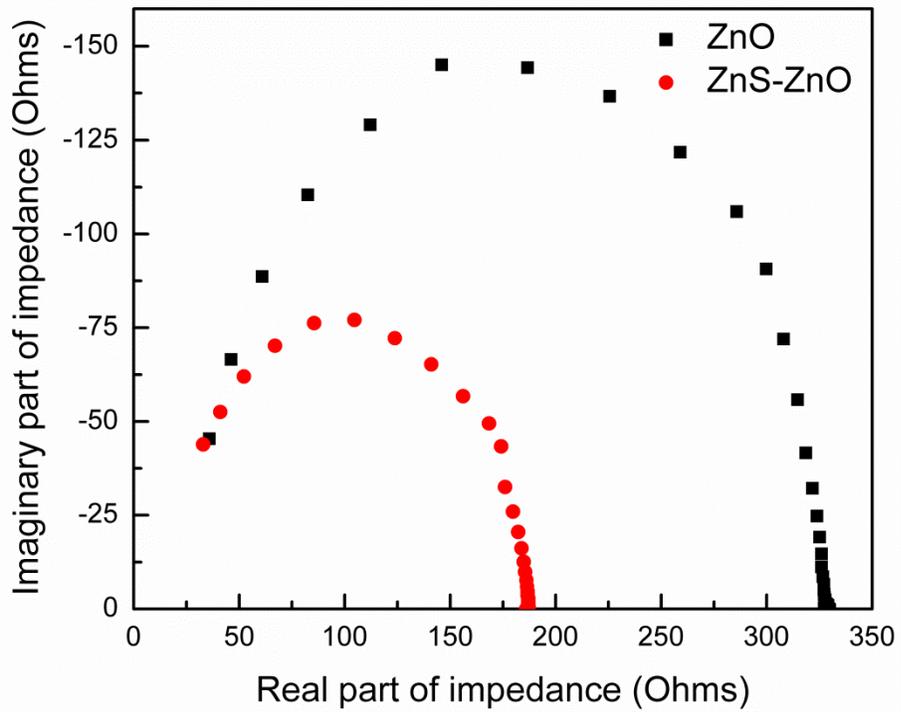

**Figure 7 (a): Impedance plane plot of ZnO nanorods and ZnS-ZnO hybrid nanostructures**

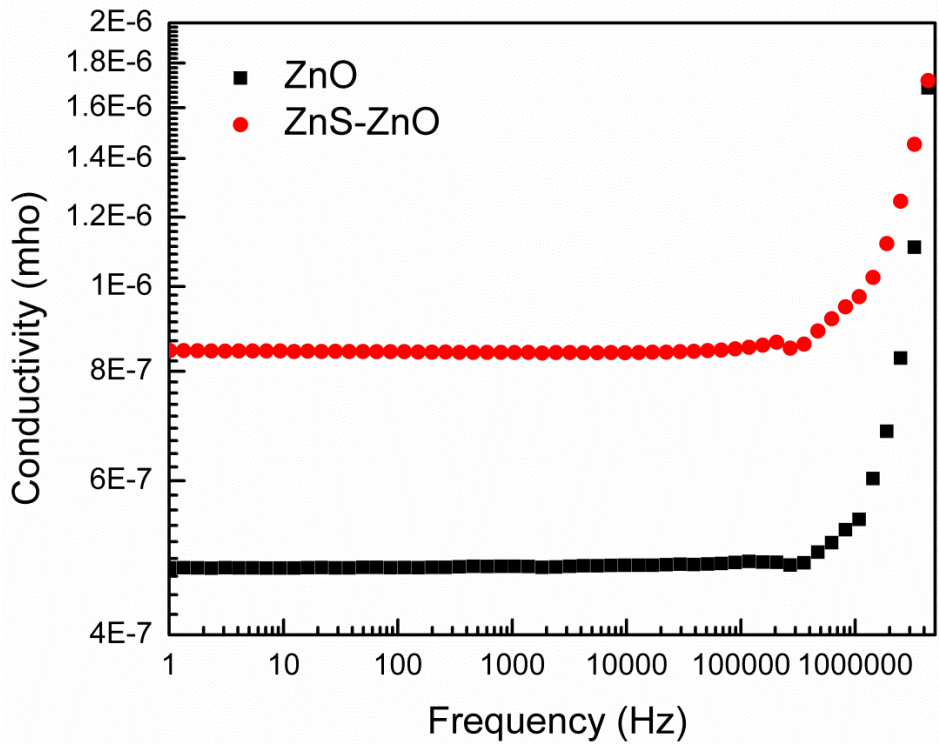

**Figure 7 (b): log-log plot of conductivity for ZnO and ZnS-ZnO hybrid nanostructures**

## 4. Electronic Structures from DFT Calculations

To address the exact source of the defect emission and have some qualitative understanding, we carried out DFT calculations using the cross section of ZnO Nanorods (NRs), which is shown in Fig. 8(f). We considered different types of intrinsic defects, i.e., oxygen vacancy $V_O$, zinc interstitial $Zn_i$, zinc vacancy $V_{Zn}$, at the surface of ZnO NRs. For $V_O$, we considered two types of O vacancies, i.e., O vacancy located in the interior of the NR (marked as 1 in Fig.(8,f)), which is coordinated by four Zn atoms, and at the edge of NR (marked as 2 in Fig.(8,f), which is coordinated by three Zn atoms. In both the cases, we relaxed all the atoms and calculated their defect formation energies $E_f$ in O-rich and Zn-rich conditions. The calculated $E_f$ of $V_O$ located in the interior of NR is 4.10 (0.93) eV in O-rich (Zn-rich) condition, whereas $V_O$ located at the edge has 2.90 (-0.27) eV in O-rich (Zn-rich) condition. On the other hand, $V_{Zn}$ at the surface of ZnO NR has 3.04 (6.20) eV formation energy in O-rich (Zn-rich). The defect formation energy of zinc interstitial $Zn_i$ is 3.20 (6.4) eV in O (Zn)-rich condition. Therefore, based on these DFT calculations, we confirm that O vacancy located at the edge of NR has the lowest formation energy. The edge O vacancy also distorts the structure and Zn-Zn bond forms with the bond length between the Zn-Zn atoms being contracted by ~0.50 Å, whereas the Zn-O bond length is expanded by ~0.05 Å. Such structural rearrangements also affect the electronic charges on the Zn and O atoms. Our detailed Mulliken charge analysis show that when an O vacancy is created at the edge of ZnO NR, the dangling bonds at the nearest Zn sites are saturated through Zn-Zn bond formation. Such bond formation also leads to a charge transfer and more charges are accumulated on the Zn atom around the O vacancy. The electronic charges of Zn atoms around the O vacancy are increased by ~0.30 e. On the other hand, no significant charge accumulation is observed at the O sites near the O vacancy.

We also calculated the band structure of relaxed pristine ZnO NR and $V_O$. Figure 8(a) shows that the pristine ZnO NR is a semiconductor, and no impurity bands can be seen in the band gap. However, a clear impurity band above the valence band can be seen in the band gap, when O vacancy at the edge of NR is introduced (see Fig.8 (b)). We believe that a $V_O$-induced state could be a possible source of defect emission in ZnO NR as observed in the PL data. Note that Zhang *et. al.*, also found a similar behavior using first-principles calculations [55]. To further diagnose the origin of O vacancy driven state in the band gap, we analyzed the charge density of all Zn and O atoms near the O vacant site close to the valence band. Fig.8 (g) shows that most of the charge is accumulated on the Zn atoms consistent with our Mulliken charge analysis. The calculated partial density of states (Fig.8(j)) further elucidates that the O vacancy-driven impurity state is mainly contributed by the Zn-4s electrons and partially by the O-atoms lying at the edge of the NR. The electronic band structure of Zn interstitial is also analyzed (see Fig. 8(d)), and one can see that the Fermi level is inside the conduction band, i.e., Zn interstitial in ZnO behaves as *n*-type impurity. One can clearly see the three Zn interstitial-driven impurity bands in the bandgap of ZnO NR, and these impurity bands can also behave as trapping center for electrons. Similarly, the Fermi level lies in the valance band of ZnO NR when Zn vacancy is introduced (Fig.8 (e)), and Zn vacancy behaves as a source of *p*-type doping. The Zn vacancy-driven impurity bands are formed just above the valence band, which indicates that electron can also transit from conduction band to these defects-driven bands.

To address the atomic origin of ZnS-ZnO reduced defect emission (increased transmittance) in the visible range, we considered placing the S atom at Zn site and at the O vacant sites 1 and 2 as marked in Fig. 8(f). S interstitial in ZnO was also considered. Our detail DFT thermodynamics show that S prefers to be doped at O vacant site 1, i.e., edge O site due to low coordination number, with formation energy of -3.98 (-7.13) eV in O-rich (Zn-rich) condition. The optimized bond length of Zn-S is about 2.30 Å and 2.22 Å, and S increases the bond length by 0.23 Å, and the Zn-O bond length is not too much affected by S. Such increment in the bond length can be expected because the atomic size of S (100 pm) is larger than the O atom (60 pm). However, as S is isoelectronic to O, it is expected that S will not bring major electronic changes in ZnO. Fig. 8(c) shows the calculated band structure of S doped at O site 1 where no impurity-driven gap state can be seen. Hence, the reduced ZnS-ZnO defect emission (increased transmittance) in the visible range can be attributed to filling the surface O vacancy by the S atom, and no electron trapping center are present in contrast to surface O vacancy band structure. We further analysed the charge density (Fig.8(h)) and the partial density of states (Fig.8(k)) at the top of the valence band, and we can see that electron density, which is mainly contributed by S-*p* and O-*p* electrons, is mainly localised around the S and edge O atoms. Fig. 8(k) clearly shows that S-*p* and O-*p* orbital are strongly hybridized.

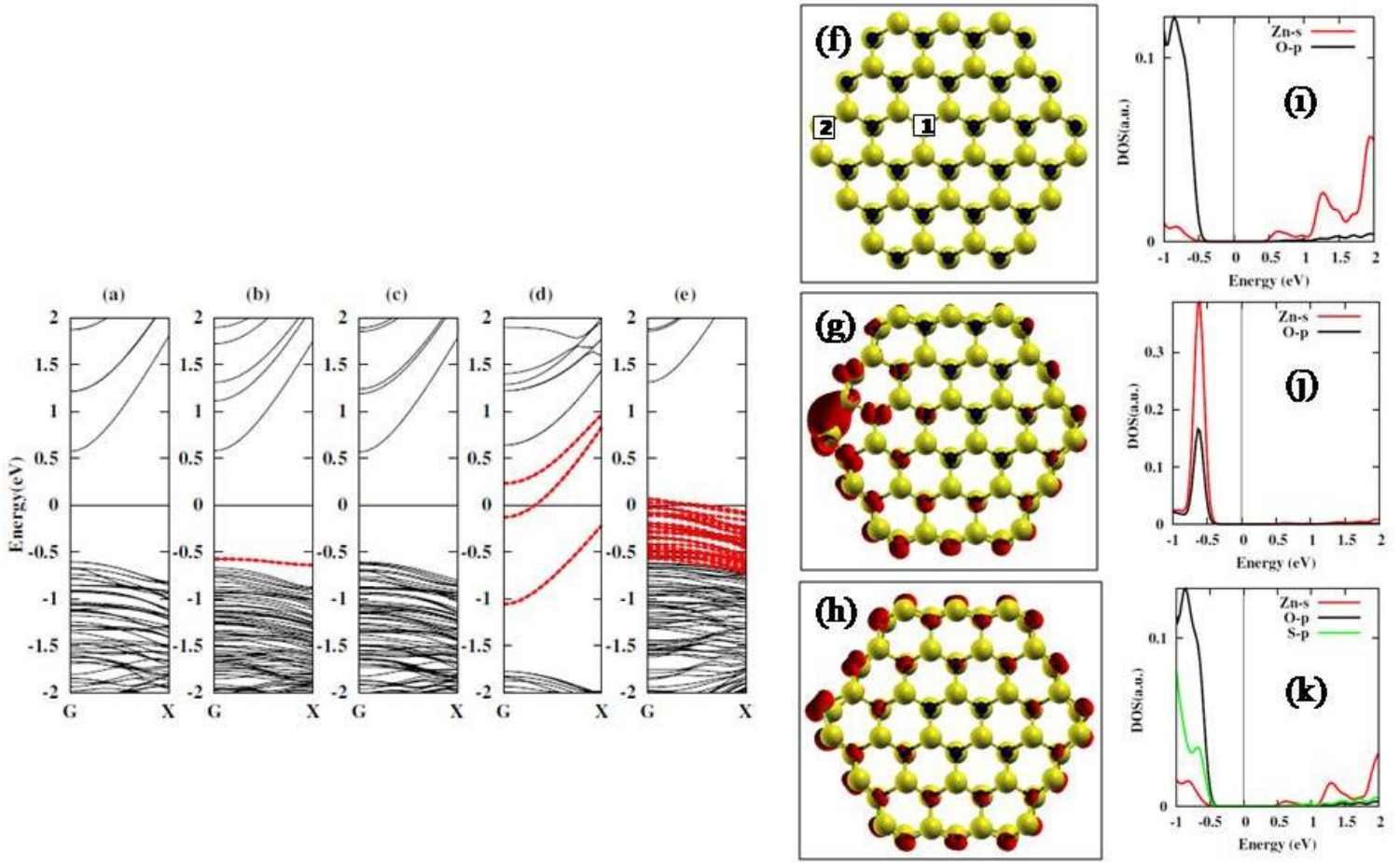

**Figure 8. DFT calculated electronic band structures of pristine ZnO NW (a), ZnO NW with oxygen vacancy (b), S substituted ZnO (c), Zn interstitial ZnO (d), Zn vacancy ZnO (e). The horizontal lines show the Fermi energy which is set at zero eV. The red bands in (b,d,e) highlight the defects driven bands. The cross-section of the top of the ZnO-NW is shown in (f), where 1 and 2 represent the oxygen surface vacancies at bulk-like and edge site, respectively. The calculated charge density of O-vacancy (g) and S substituted at O site (h). The atomic project partial density of states of pristine (i), O-vacancy (j), and S substituted at O site (k) are also shown where the vertical lines show the Fermi energy which is set at zero eV.**

## 5. Conclusion

In this study we have successfully grown ZnO nanorods and ZnS-ZnO hybrid nanostructures, and investigated the effect of ZnS surface layer on the photoluminescence and electrical properties of ZnO nanorods. The PL spectrum of ZnO nanorods showed two emitting bands i:e, band edge emission and the defect emission in the visible range. The observed defect emission has been associated with different types of defects such as $V_o$, $Zn_i$, $O_{zn}$, $V_{zn}$ and $O_i$. After growing

ZnS surface layer on ZnO nanorods, we observed a significant decrease in the defect emission of the nanorods while the band edge emission was enhanced. We conclude that the reason for the reduced defect emission in hybrid ZnS-ZnO system is associated with the doping of S atoms at the vacant oxygen sites. Our DFT calculations have supported this fact and show that the filling of oxygen vacancies with the S atoms results in the disappearance of the mid-gap states. The other oxygen related defects such as $O_i$ and $O_{zn}$, may have been reduced during the sulfurization process of ZnO nanorods. As discussed in the text the growth process of ZnS involves the dissolving of lattice O of ZnO and replacing it with the S atoms. We conclude that some of the $O_i$ and $O_{zn}$ may have been dissolved during the sulfurization process and as a result we obtained a reduction in the emission associated with these defects. The $Zn_i$ defects may have been reduced by making bond with the S atoms during the sulfurization process of the nanorods to form ZnS which is the most preferable stable state possible. Hence sulfurization of ZnO nanostructures appears as a promising method to improve the optical properties of ZnO leading to better performance of ZnO bsed opto-eletronic devices.

## 6. Acknowledgments

AA thank Kamran Amin; National Center for Nanoscience and Technology (NCNST) Beijing China for the SEM measurements of the samples. MS acknowledges the DAAD Germany for partial financial support through German-Pakistani Exchange Project "I-mod.## 7. References